\documentclass[12pt]{article}
\usepackage{graphicx}  
\begin{document}

\begin{center}

{\Large Two-dimensional PIC-MCC 
simulations of capacitively coupled radio-frequency discharge in methane}

A L Alexandrov
\footnote{To whom correspondence should be addressed (a\_alex@itam.nsc.ru)}
and I V Schweigert  

Institute of Theoretical and Applied Mechanics, 
Novosibirsk, 630090, Russia

\end{center}

\begin{abstract}
Two--dimensional capacitively coupled radio frequency discharge in methane
is simulated by PIC-MCC method.
The results were obtained in pressure range 50-300 mTorr and 
voltage range 40-180 V for discharge frequency 13.56 MHz. The electron
energy and electron--methane reaction rates spatial distributions show
existence of two regimes of discharge glow: a) with active sheaths, when
electrons are hot in electrode sheaths and cold in the middle of discharge so 
the electron--neutral reactions strongly dominate 
in sheaths regions; b) with volume domination, when the electron
energy is more uniform and the reactions take place in all discharge volume.
The second regime is usually observed for low discharge voltages, and turns 
to the first one with voltage increasing. Besides, simulation of chemical 
reactions in methane plasma was also fulfilled to find the gas mixture 
composition in discharge volume. The results are in agreement with the 
known experimental data.
\end{abstract}


\section{Introduction}
The radio-frequency (rf) methane plasma, used for producing carbon films in
plasma-enhanced chemical deposition (PECVD) reactors, is an object of interest 
for investigations. The numerical modelling of capacitively coupled 
radio-frequency (ccrf) plasma reactors has a great 
importance for understanding the processes in methane plasma and their 
influence on carbon film deposition, which helps for reactor design and 
improvement of plasma technologies.

Basic experimental information about CH$_4$ ccrf discharge plasma composition 
was obtained by group of Sugai \cite{Sug1,Sug2}.
The numerical models of methane plasma have been intensively developed 
during the recent decade. The most widely used approach is the fluid 
plasma model (\cite{Tachibana}--\cite{Herrenb2}). 
A comprehensive overview of their results
can be found in \cite{Bera1} and \cite{Bera}. 
In most of these works, the electron 
energy distribution function (EEDF) is found using various approaches and then
the rate constants of electron-neutral reactions are evaluated by integration 
of EEDF with the known energy-dependent cross section of each reaction.

Another technique, more accurate but demanding much more computation efforts,
is direct particle simulation \cite{Birdsall}, called particles-in-cells 
Monte-Carlo collision method (PIC-MCC). This approach allows to obtain EEDF
and the rates of electron-neutral reaction by direct Monte-Carlo simulations 
of particles trajectories. For methane discharge, such simulations 
in one-dimensional case were performed in \cite{Nagayama} and \cite{Rakh}. 
In work \cite{Nagayama}, each type of chemical species present in plasma was
treated by particle simulation method, but to make calculations efficient, a 
limited number of species (electrons, ions and 5 neutrals) was chosen, 
and the scheme of chemical reactions was simplified. 
Besides, the electron-impact vibrational excitation reactions, which affect the 
electron energy distribution, were not considered. Another 
approach is presented in 
work \cite{Rakh}, where the PIC-MCC simulations were proceeded
only for electrons with accounting of 18 electron--neutral reactions, 
thus providing the reaction rates, while 
the kinetics of ions and neutral species (with total number of 20)
was treated using diffusion--drift approximation and mass balance.

Since pioneering work of Levitskii \cite {Levitskii} 
the different modes of ccrf discharge operation were 
studied intensively in the experiments (\cite 
{Godyak1992}--\cite{Andujar}) and 
numerically (\cite {Parker1992}--\cite{Boeuf1992}). 
Godyak  \cite {Godyak1992} have studied 
experimentally the transition between 
the low voltage and high voltage modes in argon and 
helium. In kinetic simulations \cite {Parker1992} and 
applying a two--electron--group fluid model \cite{Boeuf}
the $\alpha - \gamma $ transition in a rf discharge 
was 
studied in helium. 
Another type of heating--mode transition was found in 
the 
experiment in 
a low pressure argon discharge \cite{Godyak1990}.  
An increase of the electron temperature 
in the midplane with pressure growth was associated in \cite {Godyak1990} 
with a change of mechanism of 
electron heating, involving the Ramsauer effect.
In the ccrf discharge in silane 
the transition between different 
modes was studied experimentally \cite{Perrin,Andujar} 
and numerically \cite{Boeuf1992}. The rise of the 
$\alpha-Si:H$ deposition rate in the volume dominated 
mode was detected in the experiment 
\cite{Perrin,Andujar}. 
In work \cite {1D} the transition between different 
modes of ccrf discharge in methane was studied in one--dimensional 
simulations with using the combined PIC--MCC 
algorithm. The phase diagram of mode location was 
constructed for wide range of gas pressures and 
discharge currents. The hysteresis was found in the 
ccrf discharge behavior with current variation. 

In this work, we present a two-dimensional (2D) PIC-MCC simulation of ccrf methane 
discharge in an axisymmetrical reactor, 
performing the PIC-MCC approach for both electron and ion kinetics. 
One of the points of interest was the existence of different regimes 
in 2D simulations. 
Besides, the gas phase chemistry 
in discharge plasma was also simulated and compared with experiment.

\section { Simulation algorithm}

\begin{figure}
\centering
\includegraphics[width=2.3in]{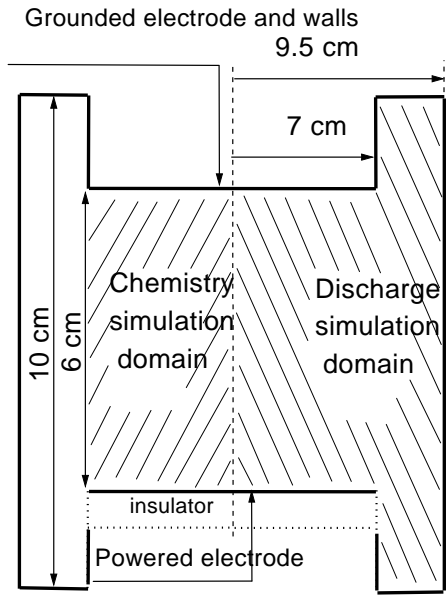}
\caption{ Geometry of reactor, showing physical domains for
plasma and gas phase chemistry simulation.}
\end{figure}

The developed PIC-MCC approach is two-dimensional in space and 
three-dimensional for particles motion (2D3V). 
In Monte-Carlo collisions simulations of electron kinetics, besides 
the elastic scattering, 
six electron-methane reactions were considered (listed in table 1), 
with the same cross sections as in \cite{Herrenb} and \cite {Rakh}.
Electron--electron Coulomb scattering, secondary electron emission and 
wall reflection were neglected so all particles reaching the chamber walls
were eliminated.

In ion kinetics, for simplicity, only one type of ion CH$^+_5$ was considered,
which is dominant in methane plasma \cite{Sug1} and other main types 
of ions present in methane (C$_2$H$^+_5$, CH$^+_4$, CH$^+_3$) have a similar 
form of density profiles (shown, for example, by simulations in \cite{Herrenb}), 
so it is possible to represent the total amount of ions by
one type only. Methane discharge plasma can be treated as 
electropositive \cite{Gogolids}, so negative ions were not taken into account.
For positive ion, the transport cross sections were taken to match the
experimental results of ion mobility in methane \cite{Urq1,Urq2}.
Ion--ion and ion--electron collisions were not considered.

The time of free flight in MCC simulations is treated by 
null collisions method.
The equations of motion were solved by explicit scheme, the time step
was chosen 10$^{-11}$ s for electrons and 40 times larger for ions.
The number of simulated particles usually was 
100000 for both electrons and ions. 
The space charge density, used for solving Poisson equation, is approximated 
using particles--in--cells technique. 

The self-bias voltage was adjusted
to keep time-averaged current equal to zero and its value was imposed on 
the grounded electrode. Potential on powered electrode was set equal
to applied rf voltage. The Poisson equation was solved 
on two-dimensional cylindrical grid with 150-200 nodes in discharge axis
direction and 50-80 nodes in radial direction, condensing near outer electrode
edge. 

The main assumption of discharge model is that the kinetics of charged particles
was simulated in pure methane. As the characteristic time of relaxation of 
gas mixture composition is of 10$^5$-10$^7$ rf cycles, for the complete PIC-MCC
discharge simulation including plasma chemistry dynamics a special algorithms 
devoted to this problem should be developed, which is the subject 
of further work. 
The calculations of plasma chemical composition using electron-methane 
reaction rates obtained in discharge simulations are presented in Section IV. 
It was shown, for methane pumping through the reactor chamber at such rates
that time of gas residence in discharge volume $\tau$ is of 0.1 sec by order 
of magnitude, the most abundant chemical species usually have 
densities 10$^3$ times less than methane (see below), so we expect the kinetics 
should not change critically. Also the abundance of excited state of methane is
neglected, which is a common assumption \cite{Nagayama,Rakh}.

The axisymmetrical physical domain for plasma simulation is shown in figure 1. 
The chosen dimensions of reactor refer to the experimental setup \cite{Sug1}. 
The gas-phase chemistry simulation domain is also shown.

\begin{figure}
\vspace{1cm}
\centering
\includegraphics[width=2.3in]{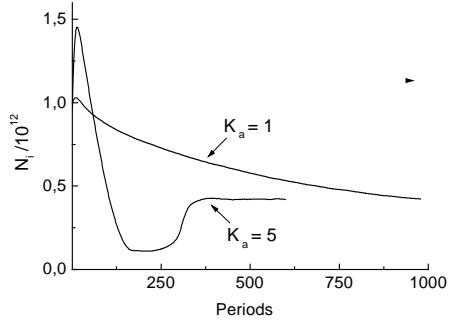}
\caption{ Relaxation of total ions number in plasma during
PIC-MCC simulation for two values of acceleration factor K$_A$.}
\end{figure}

\begin {tabular}[t]{clc} 
\hline
\hspace*{0.5cm} & Reaction       &          Energy threshold, eV    \\
\hline
& Vibrational excitation & \\
1 \hspace*{0.5cm}&  CH$_4$ + e = CH$_4^*$ + e  &  0.162     \\
2 \hspace*{0.5cm}&  CH$_4$ + e = CH$_4^*$ + e  &  0.361     \\
\hline
& Dissociation & \\
3 \hspace*{0.5cm}&  CH$_4$ + e = CH$_3$ + H + e  &   8.0    \\
4 \hspace*{0.5cm}&  CH$_4$ + e = CH$_2$ + 2H + e  &  8.0    \\
\hline
& Ionization & \\
5 \hspace*{0.5cm}&  CH$_4$ + e = CH$_4^+$ + 2e   &  12.6     \\
6 \hspace*{0.5cm}&  CH$_4$ + e = CH$_3^+$ + H + 2e &  14.3   \\
\hline
\end{tabular}

\vspace{0.5cm}

Table 1.
Electron-methane reactions involved in Monte Carlo collisions simulation.
\vspace{1cm}

To accelerate the convergence of PIC-MCC simulation,
the algorithm described in \cite{Schw} was applied.
The advantage of this method 
is illustrated in figure 2, where total number of ions in discharge 
volume during simulation is plotted versus number of simulated rf periods for 
two different values of acceleration factor K$_A$ (explained in \cite{Schw}).
The simulations were proceded for geometry shown in figure 1,
pressure 50 mTorr and $U_{rf}$=80 V. It can be seen,
that the convergence of simulation results is by nearly K$_A$ times faster,
thus saving the computational amount by the same factor. 
In practice, K$_A$ factor is limited (usually to 5-7) because it 
strongly increases the statistical noise and oscillations of solution. 
Convergence usually required 500-3000 rf cycles, the larger value belongs 
to higher voltages and pressures. Simulation of 1000 rf cycles took about 
40-60 hours on Pentium III 800 MHz.

\section { Results of discharge simulations}

Simulations were performed for discharge frequency 13.56 MHz in a 
cylindrical reactor with 14 cm diameter and 6 cm interelectrode spacing 
(figure 1) in pressure P range from 50 to 300 mTorr and rf voltage $U_{rf}$ 
from 40 to 180 V. 

The main information obtained was 2D distribution of plasma density and 
potential, mean electron energy and electron - methane reactions rates.
Previous 1D simulations of rf methane plasma \cite{Rakh,1D}
show that for some discharge parameters in the center of discharge gap, 
where electrons are trapped by the
ambipolar electric field, accumulation of relatively cold 
electrons occurs. Hence the average electron energy in this region is 
much smaller than in electrode sheaths. 
In noble gases, this effect was also observed \cite{Godyk,Tsen}.
In 1D simulations, electron energy in the sheaths may be several eV, and 
in center of discharge it may be one order of magnitude less.
Hence, the electron - methane reactions strongly dominate in the 
sheaths region.  
We call this "active-sheaths" (AS) discharge regime, in contrast to 
the another observed "volume-dominated" (VD) regime, when electron 
energy has more uniform profile and electron - neutral reactions proceed in all 
reactor volume \cite{1D}. 

We observed the existence of these regimes also in 2D simulations and 
studied the transition between them for different discharge parameters.
As an example, results of simulation
for P=123 mTorr and $U_{rf}$ = 120 V are presented in figure 3.
The electron density, averaged over period 
(figure 3(a,d)), is asymmetrical with maximum shifted to powered electrode 
due to self--bias voltage
(this was also observed in experiment \cite{Sug1} and predicted by 1D PIC-MCC 
simulation with artificial bias imposing in \cite{Nagayama}). 
It is seen that electrons in the center of discharge gap are relatively cold
(average energy is 0.75 eV while in sheaths it exceeds 2 eV), see figure 3(b) and 
also solid line in figure 3(e).
This explains the behaviour of
electrons reaction rates, which has steep maxima in regions with hot electrons.
In figure 3(c) the 2D distributon of dissociation reaction rate 
CH$_4+e\ \rightarrow$ CH$_3$+H+$e$
is shown. It is chosen for illustration as the main
methane dissociation channel; the profiles of other dissociation and ionization 
rates have a similar form. It is seen, that 
the reaction proceeds in sheath regions and also at the outer reactor edge,
and is suppressed (the rate falls by 2 orders of magnitude) in the most of 
discharge volume (see figure 3(c)). 
On the axial profile (solid curve in figure 3(f)), the absolute maximum 
of rate is located where maximum of electron energy coincides
with large electron density; a second small maximum of rate corresponds
to second maximum of energy but with lower plasma density.
For 2D case, the difference of electron energy in center and in sheaths
is not so large as in 1D simulations \cite{Rakh,1D}, where it reaches
one order of magnitude.
This may be explained by influence of reactor edge region, where 
cold electrons are not trapped by electric field so strongly as in the 1D 
discharge geometry, and hence they may diffuse in radial direction,
which leads to more flat spatial profile of electron energy. 

The described above case is an example of AS regime. 
Another discharge regime is  shown in figure 4, obtained by simulation at
P=50 mTorr and $U_{rf}$=60 V. 
Here we see a smooth profile of electron 
energy (figures 4(b) and 4(e)) and the reaction rate follows the electron density 
(see figure 4(c),4(f)). This corresponds to VD regime of discharge.

\begin{figure}
\centering
\includegraphics[width=4.3in]{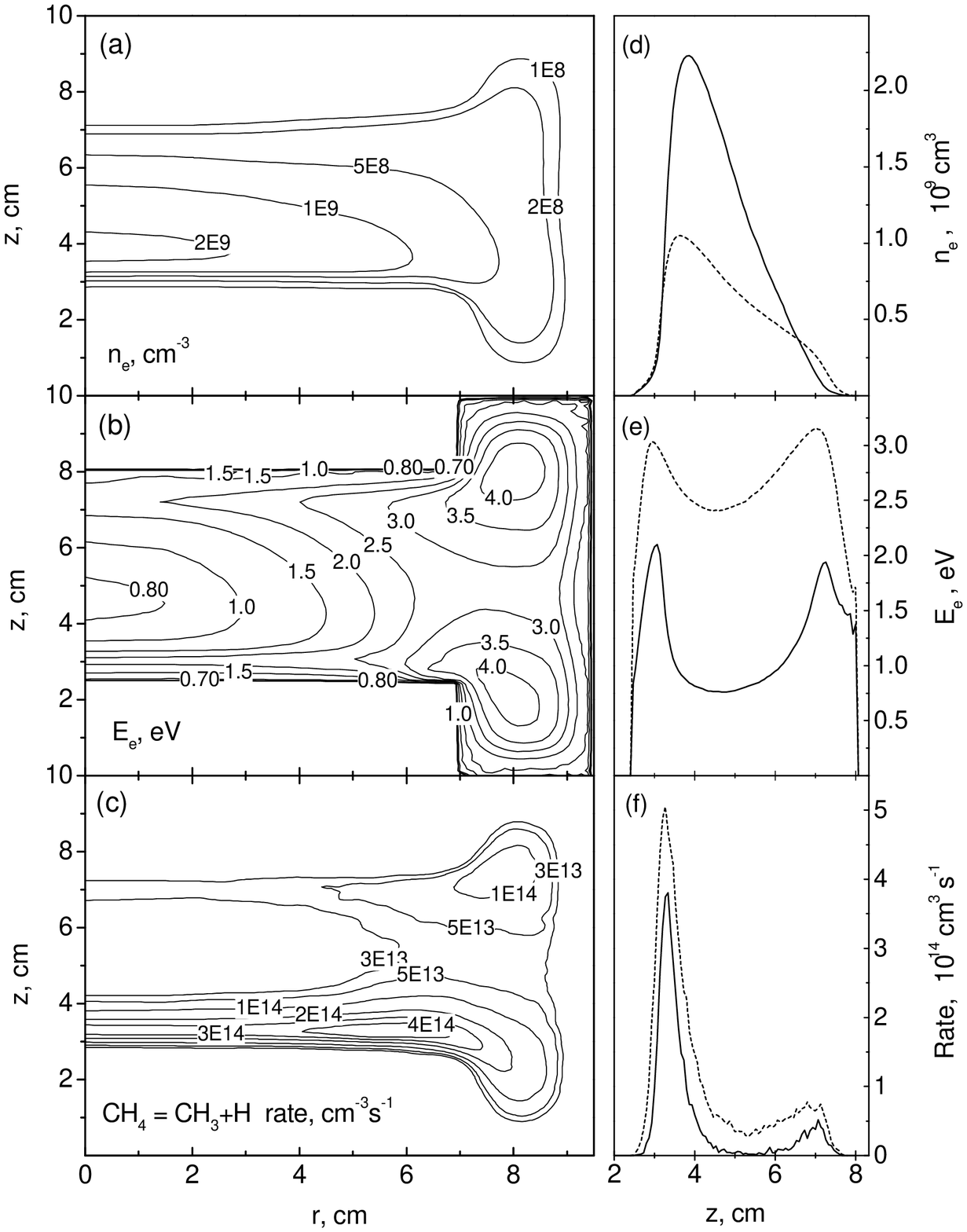}
\vspace*{-2cm}
\caption{ Contour lines for period-averaged 2D distribution of 
mean electron energy, eV (a), electron density, $cm^{-3}$ (b) 
and CH$_4 \rightarrow$CH$_3$+H dissociation rate, $cm^{-3}s^{-1}$ (c) 
for P=123 mTorr and $U_{rf}$=120 V. The 1D profiles (d-f) are plotted along the 
reactor axial direction at r=0 (solid curves) and r=6 cm
(dashed curves).}
\end{figure}

\begin{figure}
\centering
\includegraphics[width=4.3in]{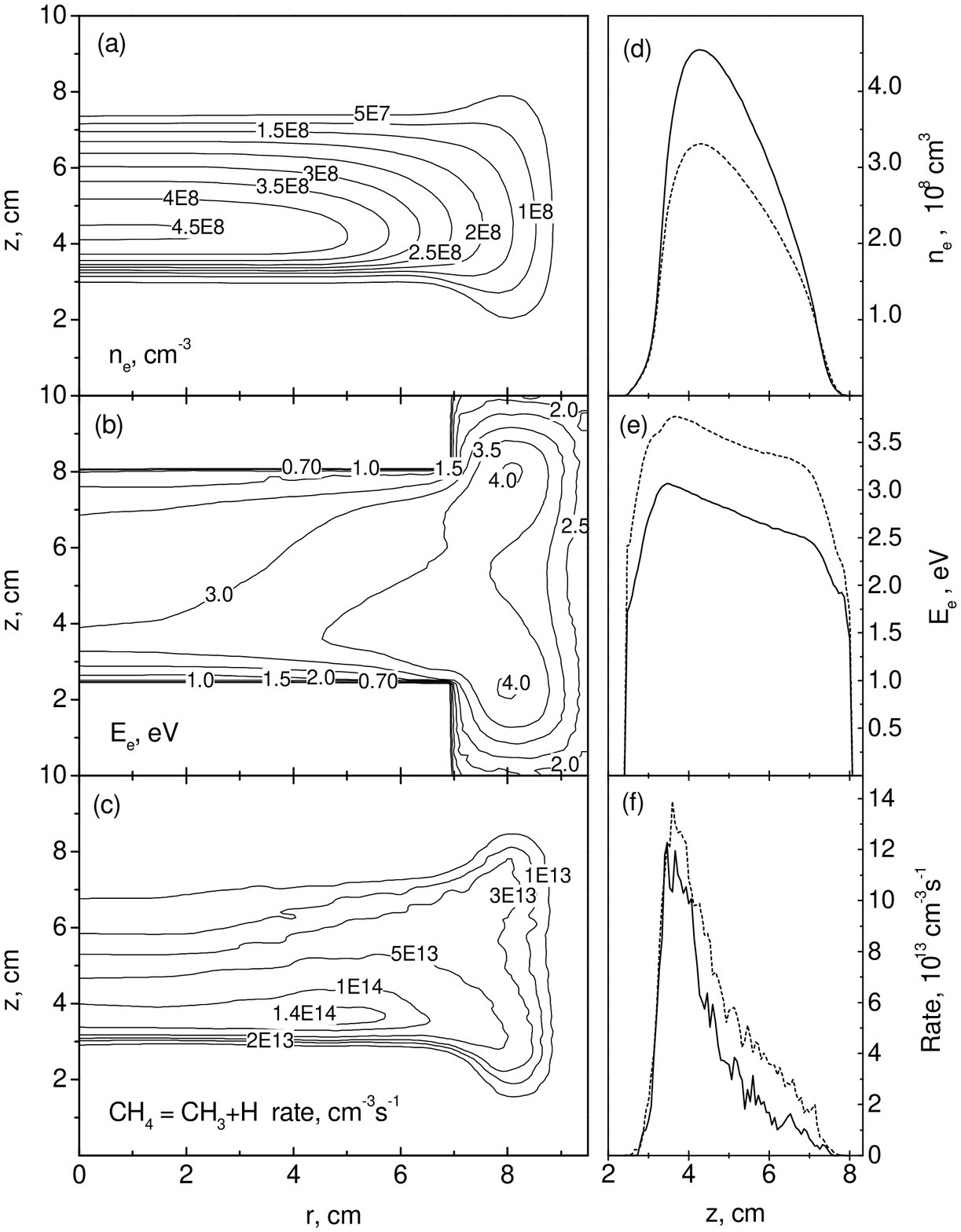}
\vspace*{-2cm}
\caption{ The same graphics as in figure 3 for P=50 mTorr, $U_{rf}$=60 V.}
\end{figure}

In general, for 2D simulations the discharge properties
are radially uniform
in the inner reactor area only (up to half of reactor radius).
In AS regime, near reactor edge the vertical electron energy profile tends 
to be more uniform (dashed curve in figure 3(e)) and 
the reaction rate is not such strongly suppressed in gap center 
(dashed curve in figure 3(f)).
Note that the average electron energy near edge is higher than in inner
region, and although the plasma density gradually decreases towards the edge,
the reaction rate even increases.

To investigate the transition between two regimes, we performed calculations 
with different rf voltages $U_{rf}$ for pressures 50, 123 and 300 mTorr.
The results are presented in 1D graphics, showing the period-averaged plasma
parameters plotted along the reactor axis (for r=0). 

For the first case,  
P=50 mTorr, the obtained plasma potential, electron density, 
mean electron energy and 
CH$_4 \rightarrow$ CH$_3$+H dissociation rate profiles 
are shown in figure 5(a-d) for several applied $U_{rf}$. 
It is seen in figure 5(a) that
the potential drop in electrode sheaths is increasing with $U_{rf}$,
and near the powered electrode (with zero time-averaged potential) 
it is few times larger than near the other one (its potential is equal to bias). 
The transition from VD to AS regime is noticeable in electron energy profiles. 
With increasing voltage, the energy decreases and for $U_{rf}$ 
over 100 V an area with relatively cold electrons
appears, making a valley on electron energy profile (figure 5(c)). A 
relatively large energy maximum appears in the sheath of powered electrode 
and a small one in the grounded sheath.
It is interesting, that increasing of $U_{rf}$ from 120 to 180 V change 
the energy profile very weak, while electron density still increase 
(figure 5(b)). The form of all density profiles is almost identical, with 
various scaling only. The maximun is slightly shifted to powered electrode
due to self--bias voltage.
The electron reactions rate profile is wide for low voltages, and narrows
with voltage increasing (figure 5(d)), indicating that the electron reactions 
tend to localize in the sheath region. 
The statistical noise is clearly seen on the rate 
profiles, despite they are averaged over 100 recent rf cycles.
To make picture readable, in figure 5(d)
not all rate profiles are shown and the curves are plotted normalized to
maximum value of each. 

\begin{figure}
\centering
\includegraphics[width=4.3in]{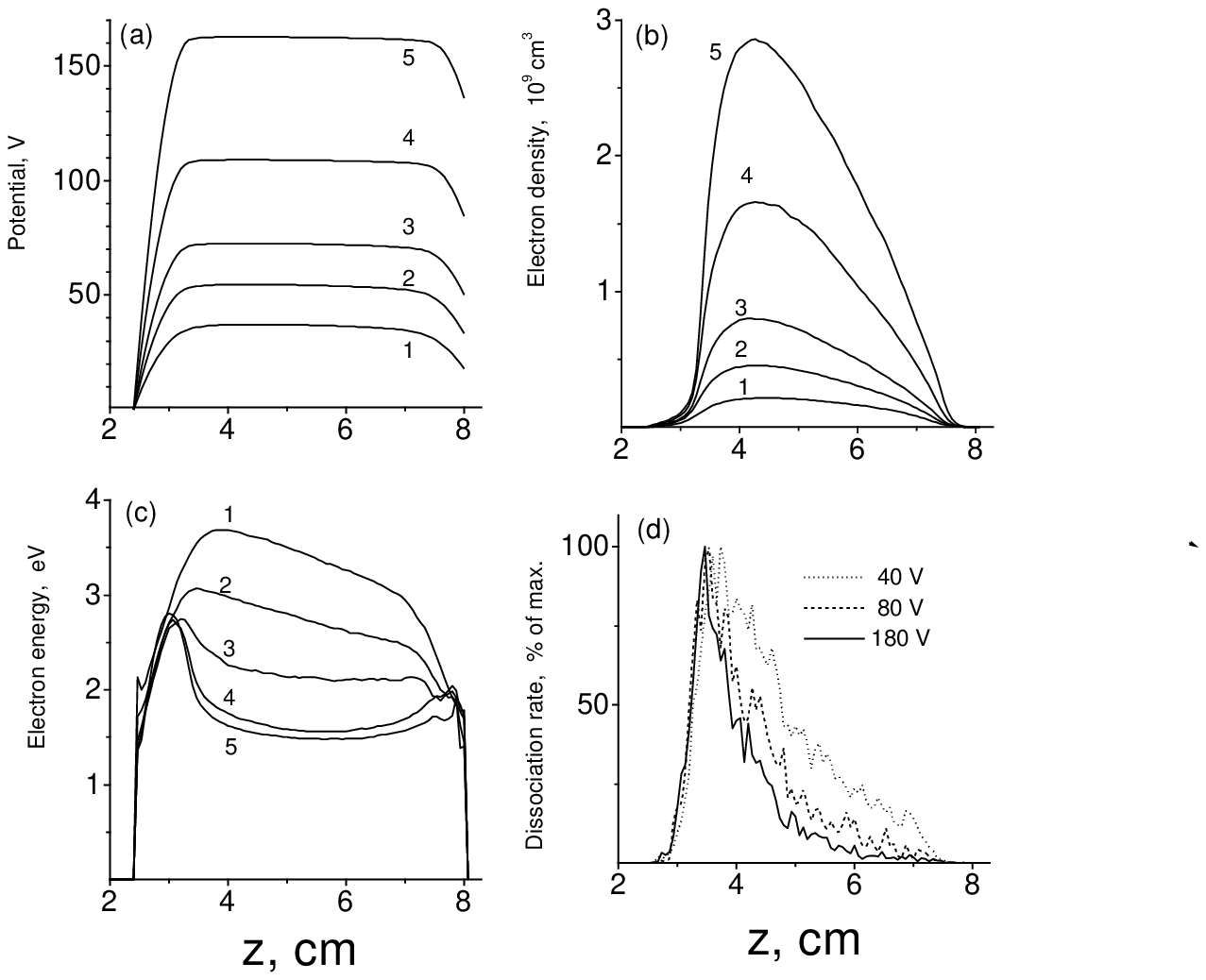}
\caption{  Axial profiles (r=0) of period-averaged plasma potential (a), 
electron density (b),
mean electron energy (c) and CH$_4 \rightarrow$CH$_3$+H dissociation rate (d)
for P=50 mTorr, 
and different $U_{rf}$: 40 (curve 1), 60 (2), 80 (3), 120 (4) and 180 V (5).
Dissociation rate (d) is plotted normalized to its maximal value,
which is  ($cm^{-3}s^{-1}$) :
$8   \times 10^{13}$ for 40 V,
$1.8 \times 10^{14}$ for 80 V,
$4.8 \times 10^{14}$ for 180 V.}
\end{figure}

The results for P=123 mTorr are shown in figure 6. The transition from VD to AS
regime is clearly seen here. For $U_{rf}$=40 V the central minimum of electron 
energy is hardly seen (curve 1 in figure 6(c)), but the regime is already 
transitional to AS, as can be seen from the reaction rate (40 V curve in figure 
6(d)) - in the sheath near powered electrode it is 2-3 times larger than in the 
volume. With $U_{rf}$ increasing, energy minimum becomes wider and deeper,
making narrow maxima in both sheaths. Again the larger maximum is observed 
in powered electrode sheath. The second maximum near grounded electrode is 
better seen than for 50 mTorr, possibly due to larger potential drop in 
this sheath at higher pressure (compare figures 5(a) and 6(a)). 
The form of electron energy 
profile also tends to saturate for $U_{rf}$ larger than 150 V, while the form
of larger maximum becomes established near $U_{rf}$=100 V. 
Electron density increases with $U_{rf}$, as shown in figure 6(b). For 150 and 
180 V it is not shown, but
has the identical form as for 120 V, scaled to maximum 3.7$\times 10^9$ $cm^{-3}$
and 5.7$\times 10^9$ $cm^{-3}$ correspondingly. The position of density maximum
shifts to the powered electrode (as observed in \cite{Sug1}, shown by points
in figure 6(b)) and for $U_{rf}$ larger than 80 V does not move. 
The dissociation rate (figure 6(d)) exhibits two maxima near both sheath regions, 
with strong rate suppression in central region with $U_{rf}$ increasing, 
thus indicating transition to the AS regime. Larger
maximum corresponds to both energy and density maxima near powered electrode,
with $U_{rf}$ increase it strongly enlarges without broadening;
the second maximum, near grounded electrode, is smaller and diminishes with
$U_{rf}$ increasing due to more cool electrons in this sheath.
The points in figures 6(b,c) are taken from measurements made for P=123 mTorr
in \cite{Sug1}.
The best fitting with our simulations is obtained for $U_{rf}$=80 V, which
is in agreement with discharge power 10 W in experiment \cite{Sug1}.

Comparison with simulations of the same discharge parameters
using fluid model \cite{Bera2} show good agreement in plasma
potential profile (within a few V) and electron density (within 20\%)
but different electron energies, which were in \cite{Bera2} 3-4 eV
in plasma bulk and near grounded electrode but increased to 11-13 eV
in powered electrode sheath.

\begin{figure}
\centering
\includegraphics[width=4.3in]{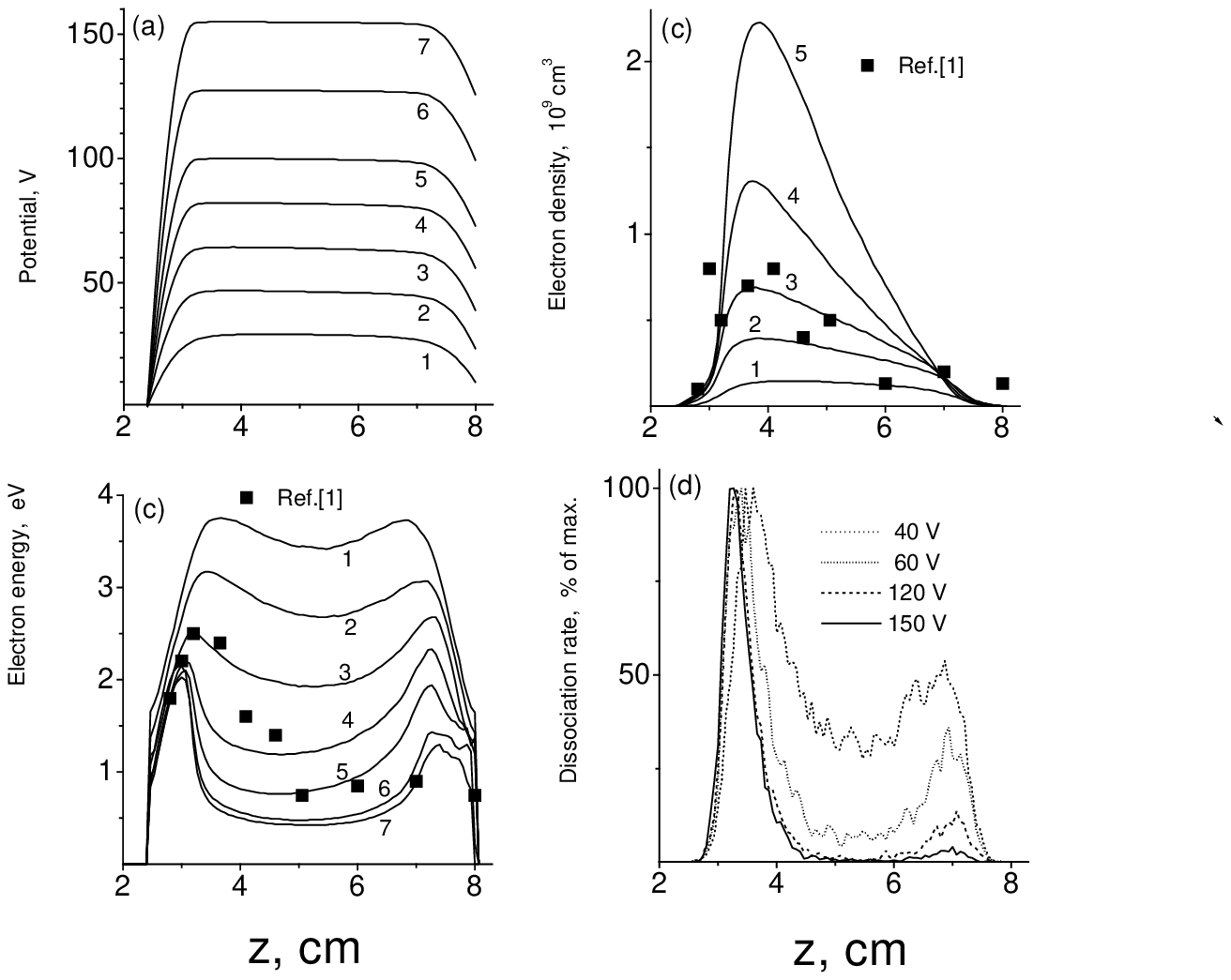}
\caption{The same graphics as in figure 5, but for P=123 mTorr and  
$U_{rf}$=40 (curve 1), 60 (2), 80 (3), 100 (4), 120 (5), 150 (6) and 180 V (7).
The points are from experiment \cite{Sug1}.
Electron densities (b) are shown not for all voltages.
Dissociation rate (d) is plotted normalized to its maximal value,
which is ($cm^{-3}s^{-1}$) :
$8   \times 10^{13}$ (40 V),
$1.8 \times 10^{14}$ (60 V),
$3.8 \times 10^{14}$ (120 V),
$6   \times 10^{14}$ (150 V).}
\end{figure}

The last set of simulations was proceeded for P=300 mTorr (figure 7). 
For $U_{rf}$=60 V the regime is VD, with symmetrical plasma density
profile and hardly noticeable minimum in energy profile (curve 1 at 
figures 7(b) and 7(c)) 
but enhanced reaction rates in sheaths (60 V curve in figure 7(d)). 
For $U_{rf}$=80 V, the regime is closer to AS (seen from the reaction rate 
profile, 80 V curve in figure 7(d)), 
the plasma density is one order of magnitude larger than 
for 60 V and asymmetrical (curve 2 in figure 7(b)), although the electron 
energy profile (curve 2 in figure 7(c)) does not exhibit a deep minimum.

With further increasing of $U_{rf}$, narrow energy maxima again appear in 
both sheaths, but unlike for the previous pressures, the maximum in 
grounded electrode sheath is larger than the other one. 
Such difference in energy profile evolution may possibly be explained
by comparing plasma potential behavior (compare figures 6(a) and 7(a)).
For 300 mTorr the potential drop 
in powered electrode sheath is by 16-18 V smaller than for 123 mTorr, 
while near the grounded one it is larger: 
for $U_{rf}$ from 80 to 180 V the drop varies from 30 to 40 V instead 
of 23 to 28 V for 123 mTorr. Sheath thickness is nearly the same for both
pressures (near 0.9 cm in AS-regime).  Like for previous pressures, the 
energy profiles tends to saturate with voltage over 150 V.
\begin{figure}
\centering
\includegraphics[width=4.3in]{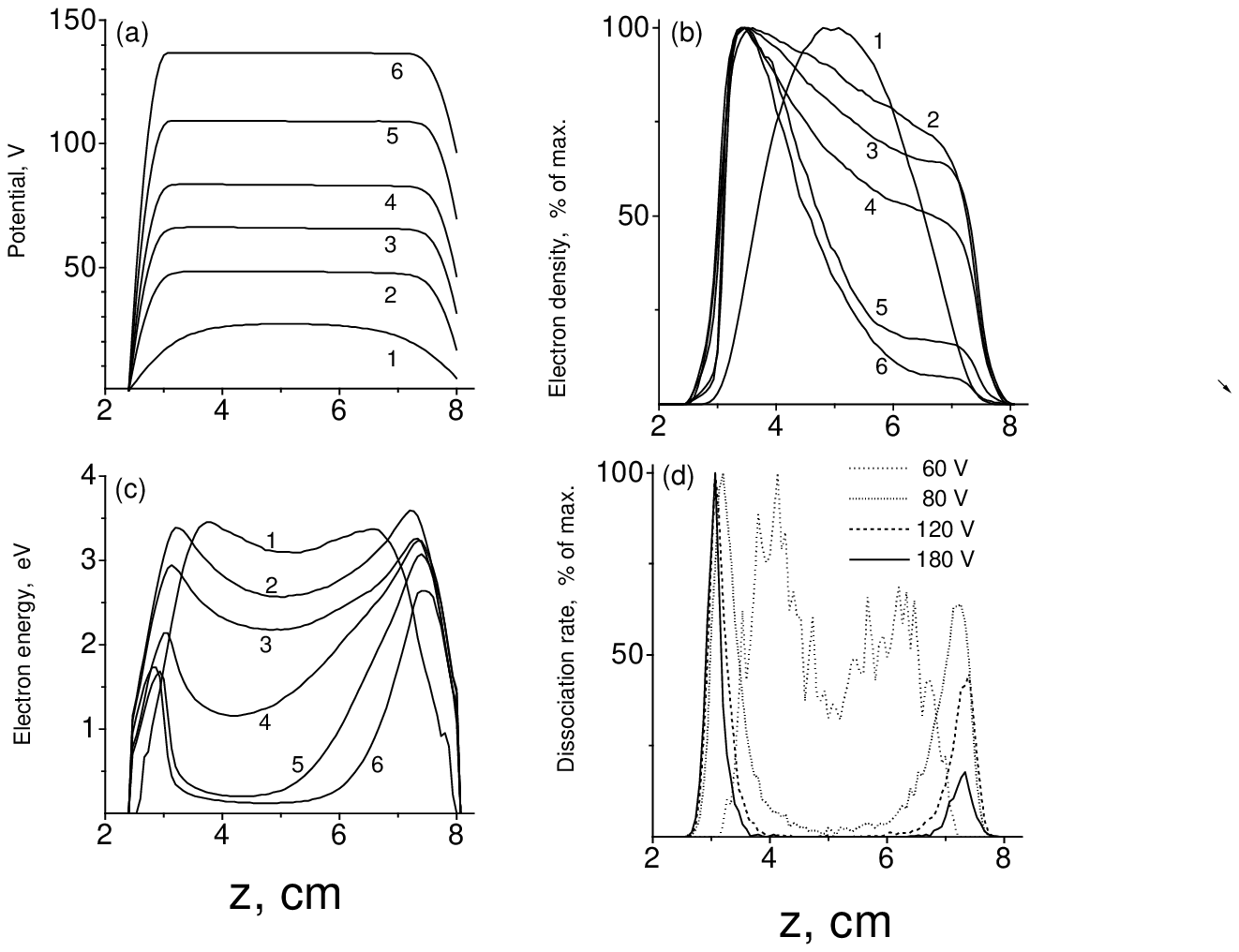}
\caption{The same graphics as in figure 5, but for P=300 mTorr and  
$U_{rf}$=60 (curve 1), 80 (2), 100 (3), 120 (4), 150 (5) and 180 V (6).
Electron densities (b) and
dissociation rates (d) are plotted normalized to maximal value.
Maximum values for densities are ($cm^{-3}$): 
$4.5 \times 10^{7}$, 
$4.2 \times 10^{8}$, 
$7.6 \times 10^{8}$, 
$1.2 \times 10^{9}$, 
$3.8 \times 10^{9}$, 
$8.5 \times 10^{9}$  
(1-6).
Maximum values of dissociation rates ($ cm^{-3}s^{-1}$): 
$1.5 \times 10^{13}$  (60 V),
$3.5 \times 10^{14}$  (80 V),
$8   \times 10^{14}$  (120 V),
$1.7 \times 10^{15}$  (180 V).}
\end{figure}

The behaviour of electron density is shown in figure 7(b). As it varies for more
than two orders of magnitude, the profiles are also plotted normalized to 
maximal 
value of each. 
However, it can be seen from this graphics that VD-AS transition 
leads to accumulation of cold electrons in ampibolar field region, where plasma 
potential has a plateau. 

\section { Gas phase chemistry simulation}

For simulation of gas phase chemistry, the balance equations for
neutral species density $n_i(r,z)$ were solved, where $r$ is the distance from 
the reactor axis (radial direction), $z$ is the distance from the powered 
electrode (axial direction). The physical domain for chemistry simulation is
reduced to interelectrode space, as shown in figure 1.
The convection is also taken into account:

$$\frac{dn_i}{dt}= R_i + div \vec J_i,$$

$$\vec J_i = D_i \vec \nabla  n_i + n_i \vec V ;$$

$R_i$ is the sum of generation and loss for i-th species due to chemical
reactions, $\vec J_i$ is the flux, where $D_i$ is diffusivity, 
$\vec V$ is the gas velocity in the plasma reactor, given as two-dimensional 
vector field.

The chemical model includes radicals H, CH, CH$_2$, CH$_3$, C$_2$H$_5$, 
and stable species H$_2$, CH$_4$, C$_2$H$_2$, C$_2$H$_4$, 
C$_2$H$_6$, C$_3$H$_8$. 
For balance equations of radicals, which have non-zero coefficient $s$ of 
sticking to the surface, the additional loss terms were included at the 
electrodes boundaries:

$$J_{ib} = \frac{1}{4}\ n_{ib} v_t \ \frac{s_i}{1-s_i/2}$$

Where $n_{ib}$ is radical density at the boundary,
$v_t$ is thermal velocity and the value of $s_i$ was assumed to be 0.01 
for all types of radicals except 0.025 for CH$_2$ \cite{Herrenb}. The 
diffusivities were taken as in \cite{Rhallabi} for radicals, and as in 
\cite{Herrenb} for stable species.

\begin {tabular}[t]{rlcc} 
\hline
N \hspace*{0.5cm} & Reaction       &                   Rate constant k$_N$, $m^3 s^{-1}$ & [Ref.]\\
\hline
 & Electron-methane  & & \\
1 \hspace*{0.5cm}&  CH$_4$ + e = CH$_3$ + H + e  &         obtained by MCC & \\
2 \hspace*{0.5cm}&  CH$_4$ + e = CH$_2$ + 2H + e  &        obtained by MCC    &   \\
3 \hspace*{0.5cm}&  CH$_4$ + e = CH$_4^+$ + 2e   &          obtained by MCC   &   \\
4 \hspace*{0.5cm}&  CH$_4$ + e = CH$_3^+$ + H + 2e &        obtained by MCC  &   \\
\hline
 & Ion-methane  & & \\
5 \hspace*{0.5cm}&  CH$_4$ + CH$_4^+$ = CH$_5^+$ + CH$_3$      &   1.5$\times$10$^{-15}$  &  \cite{Tachibana}\\
6 \hspace*{0.5cm}&  CH$_4$ + CH$_3^+$ = C$_2$H$_5^+$ + H$_2$   &   1.2$\times$10$^{-15}$  &  \cite{Tachibana}\\
\hline
 & Radical reactions & & \\
7 \hspace*{0.5cm}&  CH$_3$ + CH$_3$ = C$_2$H$_6$         & 8  $\times$10$^{-17}$ &   \cite{Bera}\\
8 \hspace*{0.5cm}&  CH$_3$ + H = CH$_4$                  & 1.38$\times$10$^{-16}$ &   \cite{Bera}\\
9 \hspace*{0.5cm}&  CH$_2$ + H = CH + H$_2$              & 2.7$\times$10$^{-16}$  &  \cite{Bera} \\
 10 \hspace*{0.5cm}& CH$_2$ + CH$_4$ = CH$_3$ + CH$_3$   & 1.5$\times$10$^{-18}$  &  fitted, see text\\
11 \hspace*{0.5cm}& CH$_2$ + CH$_2$ = C$_2$H$_2$ + H$_2$ & 5.3$\times$10$^{-17}$  &  \cite{Bera}\\
12 \hspace*{0.5cm}& CH$_2$ + CH$_3$ = C$_2$H$_4$ + H    &  10$^{-16}$    &    \cite{Bera}\\
13 \hspace*{0.5cm}& CH + CH$_4$ = C$_2$H$_4$ + H        &  10$^{-16}$    &    \cite{Herrenb}\\
14 \hspace*{0.5cm}& CH + CH$_4$ = C$_2$H$_5$            &  10$^{-16}$    &    \cite{Herrenb}\\     
15 \hspace*{0.5cm}& C$_2$H$_5$ + H = CH$_3$ + CH$_3$    &  6$\times$10$^{-17}$  &  \cite{Herrenb}\\
16 \hspace*{0.5cm}& C$_2$H$_5$ + H = C$_2$H$_4$ + H$_2$ &  3$\times$10$^{-18}$  &  \cite{Herrenb}\\
17 \hspace*{0.5cm}& C$_2$H$_5$ + CH$_3$ = C$_3$H$_8$    &  4.2$\times$10$^{-18}$ &   \cite{Herrenb}\\
\hline
\end{tabular}
\vspace{0.5cm}

Table 2. 
Chemical reactions taken into account in gas phase chemistry model.
\vspace{1cm}

\begin{figure}
\centering
\includegraphics[width=5.3in]{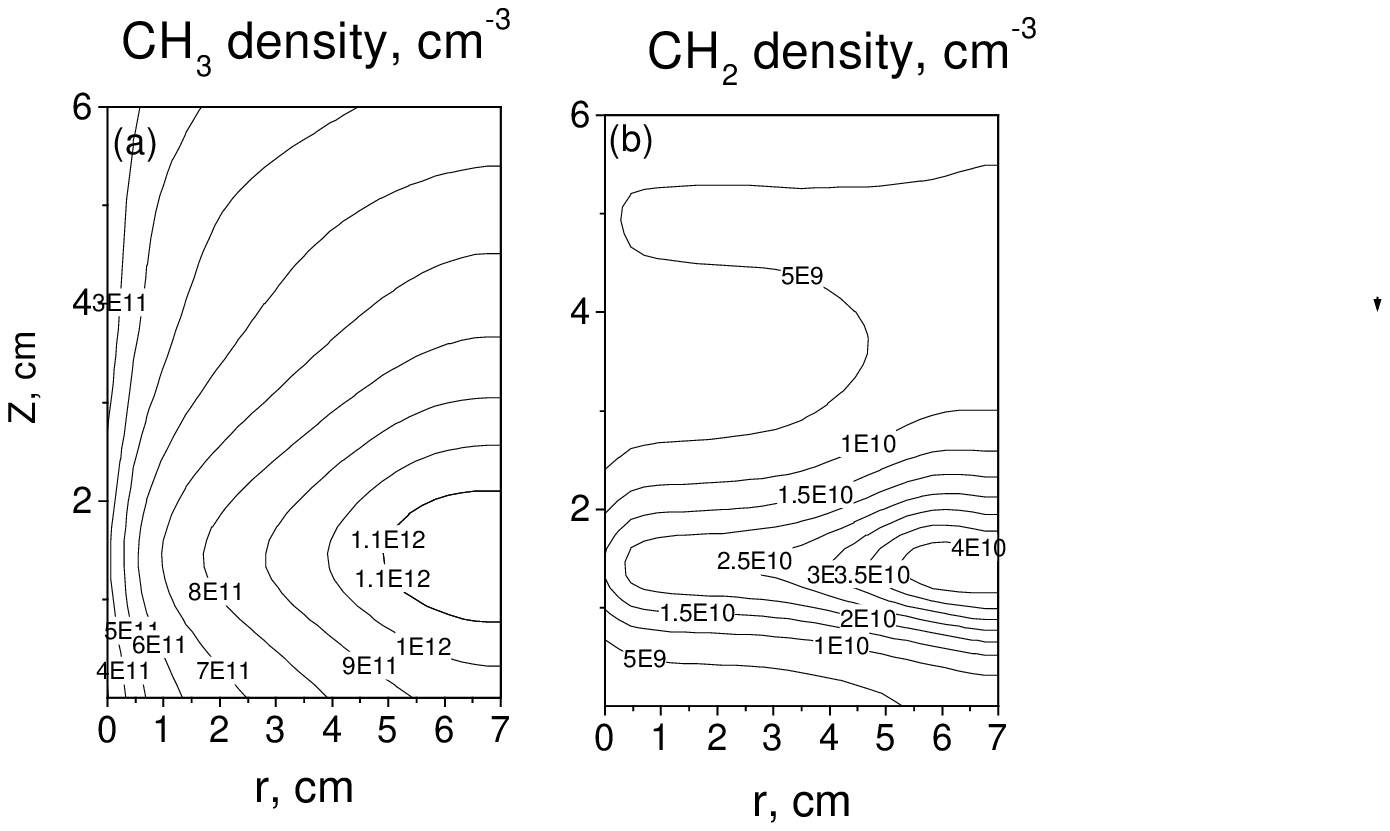}
\caption{ Calculated 2D density profiles for radicals CH$_3$ (a)
and CH$_2$ (b) obtained for P=123 mTorr, $U_{rf}$= 80 V, $\tau$=150 ms.}
\end{figure}


\begin{figure}
\centering
\includegraphics[width=5.3in]{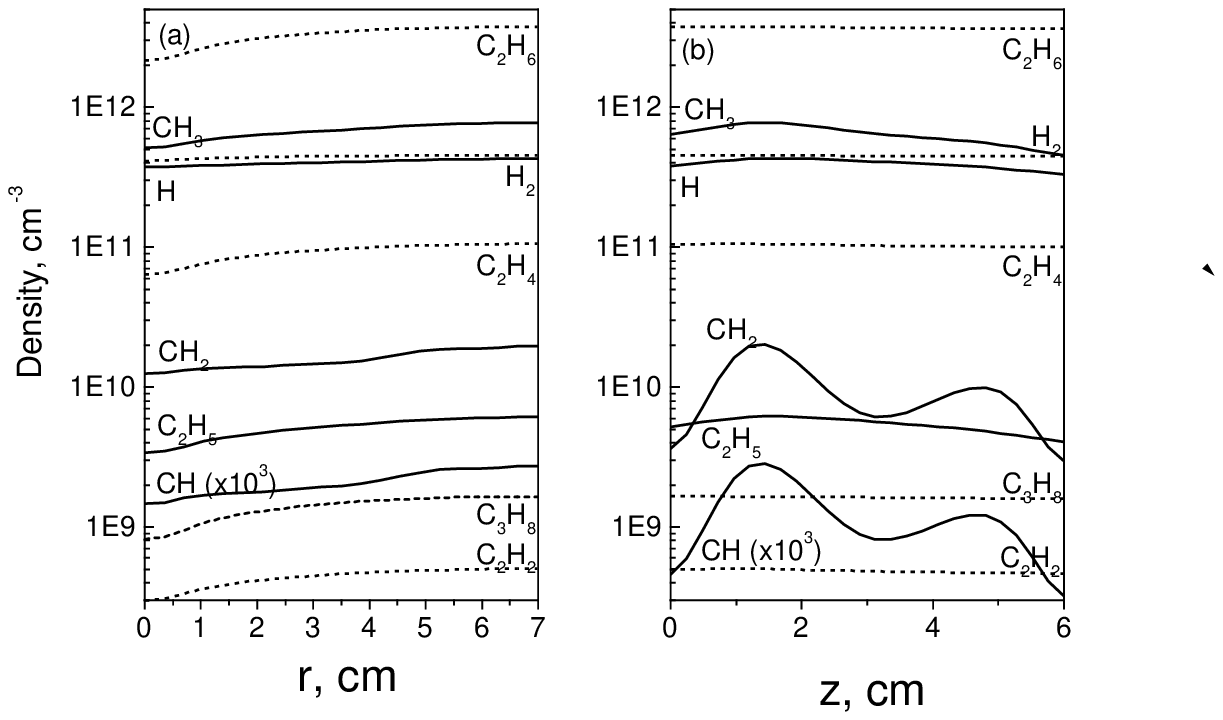}
\caption{Calculated 1D density profiles for radicals (solid
lines) and stable species (dashed lines), in radial direction  
at $z$=1.2 cm (a), and in axial direction at $r$=7 cm (b),
for the same discharge conditions as in figure 8. 
Densitiy of CH radical is multiplied by 1000.}
\end{figure}

The set of gas phase chemical reactions was taken as in \cite{Bera} with 
reactions responsible for C$_2$H$_5$ radical balance added from \cite{Herrenb} 
(see table 2).
The volume rates of electron - methane reactions are obtained in discharge
simulation as two-dimensional profiles and 
used as production terms for corresponding species. 
The account of ion - methane reactions 5 and 6 requires the information 
about ion 
mixture composition, which is not considered in PIC-MCC simulations.
However, preliminary calculations show that
a simplification can be made to exclude the ion reactions
from the chemical model.
Since the rate constant of reactions 5 and 6 are
high enough, estimation shows that CH$_4^+$ and CH$_3^+$ ions exist 
for a few free
paths only before conversion to CH$_5^+$ and C$_2$H$_5^+$, which do not take 
part in 
gas phase chemistry. As each type of ion has the only one way of chemical 
conversion, the volume rate of reaction 5 can be assumed equal to those of 3.
The same applies to reactions 6 and 4. So the reactions 3,5 and 4,6 may be 
combined and replaced by:

3,5:\hspace*{1.5cm}   2CH$_4$ + e = CH$_5^+$ + CH$_3$ + 2e

4,6:\hspace*{1.5cm}   2CH$_4$ + e = C$_2$H$_5^+$ + H$_2$ + H + 2e

with the rates equal for those of reactions 3 and 4, respectively. 
Thus the effect of ion - methane reactions is accounted through
their production rates and only neutral species remain in the chemical 
reactions system.

The rate constant of reaction 10, found in literature, ranged
from $k_{10}$= 10$^{-20}$ \cite{Sug1} to 10$^{-18}$ 
\cite{Bera,Gogolids} 
and 1.7$\times$10$^{-17}$ $m^3s^{-1}$ \cite{Tachibana}.
This rate constant strongly affects the solution for CH$_2$ radical,
because CH$_4$ has the largest density and reaction 10 is the main
loss term for CH$_2$. For the other species its influence is weak.
We used it as a fitting parameter, and
found that the best agreement with experiment \cite{Sug2} is achieved  
for $k_{10}$= 1.5$\times$10$^{-18}$ $m^3s^{-1}$, which is close to
used in \cite{Bera} or \cite{Gogolids}. 

To take into account the effect of convection, we considered reactor with gas 
inlet through the centre of reactor and outpumping at the outer
boundary.
For these model calculations the field of gas velocity $\vec V(r,z)$ 
was simplified:

$$V_z(r,z)=0$$
$$V_r(r,z)=V_0 r_0 / r,\  r > r_0\ ;\ \  V_0 r / r_0, \  r < r_0$$

where $r_0$ is the gas inlet radius (taken as 0.5 cm). Density of CH$_4$ 
was held 
constant at $r < r_0$ in order to make a feed term in convection. 
The value of $V_0$ was defined to give a chosen time $\tau$ 
of gas residence in reactor, which was varied in calculations 
from $\tau$ = 15 ms to 300 ms.
Although assuming $V_z$ to zero may be a rough approximation, especially for the
central region, calculations show that the density profiles are not very
sensitive to details of $\vec V$ field in central region, but depend mostly on 
$\tau$.

The balance equations are approximated using finite-difference scheme in
cylindrical physical domain (see figure 1) and integrated on time by 
Runge-Kutta method until the solution converged to steady state. The densities 
of radicals converge fast, while solution for stables requires physical time
of about 2$\tau$.

A typical result of two-dimensional gas phase chemistry simulation is presented
in figures 8 and 9. The electron - methane reaction rates were taken for
discharge at P=123 mTorr, $U_{rf}$= 80 V, $\tau$ was chosen as 150 ms.

Figure 8 show two-dimensional density profiles of the main radicals CH$_3$ and
CH$_2$. As the discharge regime is close to AS, the density has maxima near
electrode sheaths, where electron-methane reactions are localized 
(see figure 6(d)).
Profile of CH$_3$ is more flat due to its larger time of chemical decay.

Figure 9(a) show the profiles $n_i(r)$ in radial direction,
plotted by solid lines for radicals, which profiles are shown at $z$=1.2 cm,
where radical density is maximal, and by dotted lines for stable species 
(their profiles are almost flat in axial direction). 
Figure 9(b) shows profiles $n_i(z)$ of same species in axial 
direction at the outer edge of electrodes ($r$=7 cm), where all densities
are maximal. The profiles of CH, which
has in our calculations density less than 10$^7$ $cm^{-3}$, are enlarged by 
factor of 1000.
For $\tau$=150 ms, the most abundant stable
C$_2$H$_6$ has maximal density 10$^3$ times less than methane 
(4.3 $\times$ 10$^{15}$ $cm^{-3}$), 
so the neglecting of chemistry influence on discharge physics is possible.

\begin{figure}
\centering
\includegraphics[width=4.0in]{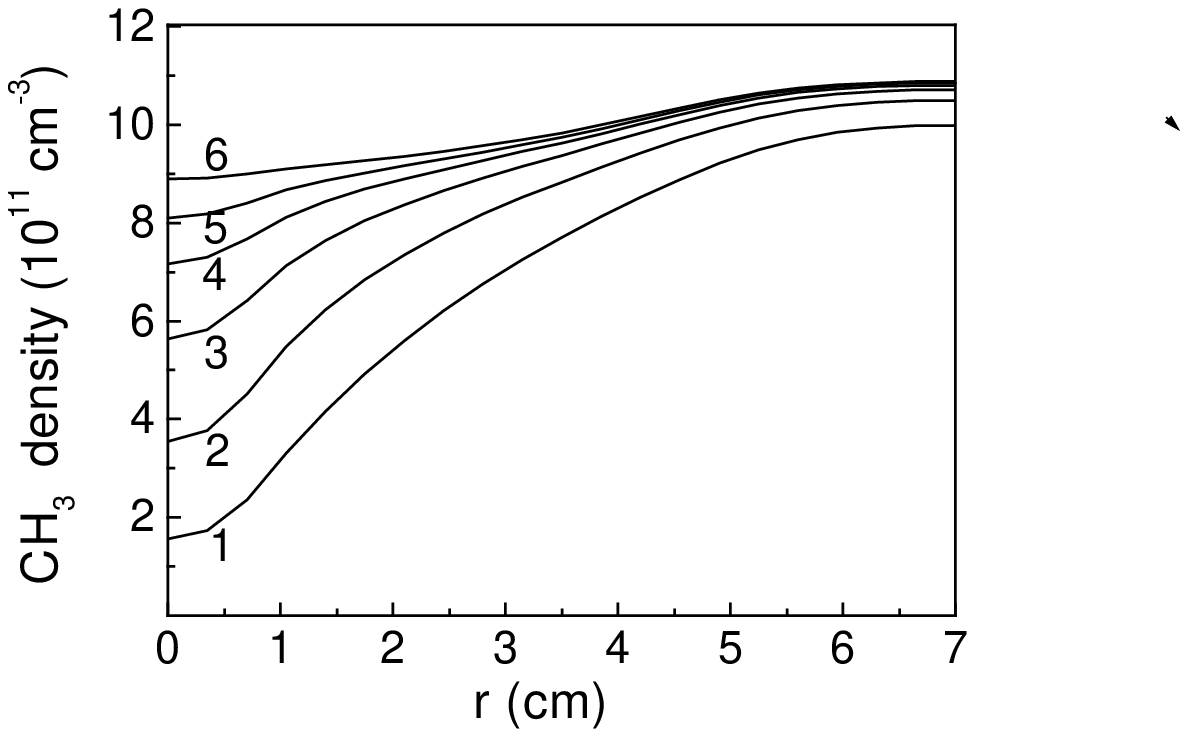}
\caption{ Calculated radial density profiles for CH$_3$ radical
at $z$=1.2 cm, for the same discharge conditions as in figure 8 but different
$\tau$: 17 ms (1), 35 (2), 70 (3), 150 (4) and 300 ms (5). Curve 6
corresponds to infinite $\tau$ (no convection).}
\end{figure}

Calculations with various $\tau$ show, that the density profiles 
of stable species, which main loss term is
convection, are approximately proportional to $\tau$.
For radicals, the profiles are determined also by diffusion and 
chemical decay, so for low gas velocities the influence of convection is weak. 
The effect of $\tau$ variation on CH$_3$ density profile is shown 
in figure 10. The upper curve was obtained with zero convection term, when the
steady state solution is achieved only for radicals.
Of course our discharge simulation remains valid only for small enough $\tau$ 
(of 100 ms order of magnitude),
while we can still neglect the change of gas composition, so this curve 
shows only the possible limit of convection influence. 
                                              
It is seen, that the density saturates with increasing of $\tau$, at first 
in the outer
region, where gas velocity is smaller. For $\tau >$100 ms it becomes saturated
for the most part of reactor, so we usually made calculations with 
$\tau$=150 ms (for our reactor geometry and P=123 mTorr this corresponds
to gas feed rate of 55 sccm, which is close to used in experiments 
\cite{Sug1,Sug2}.
The behaviour of H and C$_2$H$_5$ profiles is similar. For CH$_2$ and CH, 
the profile is 
much less sensitive to $\tau$, because the main loss term for them is the 
chemical decay.

\section { Comparison with experiment}

Calculations with various pressure were made to compare the profiles
of CH$_3$ and CH$_2$ radicals with the observed in \cite{Sug2}.
The simulated pressures were 300, 140, and 57 mTorr, the other 
conditions were the same as for figure 8. The results are presented in figure 11(a,b)
for CH$_3$ and figure 11(c,d) for CH$_2$.
Figure 11(a) shows CH$_3$ profiles, plotted at $r$= 4.5 cm, where they were measured
in experiment \cite{Sug2}, for pressures 140 and 300 mTorr. The agreement
with experiment for P=140 mTorr is fine. 
Note that the only fitting parameter was the rate constant $k_{10}$.
For P=300 mTorr the profile
exhibits two maxima, similar to experiment, but the calculated 
density is some larger, especially near the grounded electrode.
Figure 11(c) shows the same for CH$_2$. The agreement for 140 mTorr
is good again, but profiles are more steep than
in experiment, especially for 300 mTorr, where calculated density
is very small in the centre of discharge gap. This may be caused by
underestimation of the electron--methane reaction rates in the discharge center
for high pressures. This shows that for pressures of 300 mTorr and higher the
developed kinetic model is not very accurate. 
In figures 11(b) and 11(d) the both radical profiles are shown for 57 mTorr.
Although the calculated profiles are more steep than the observed, 
the agreement can be considered as satisfactory. 

\begin{figure}
\centering
\includegraphics[width=5.3in]{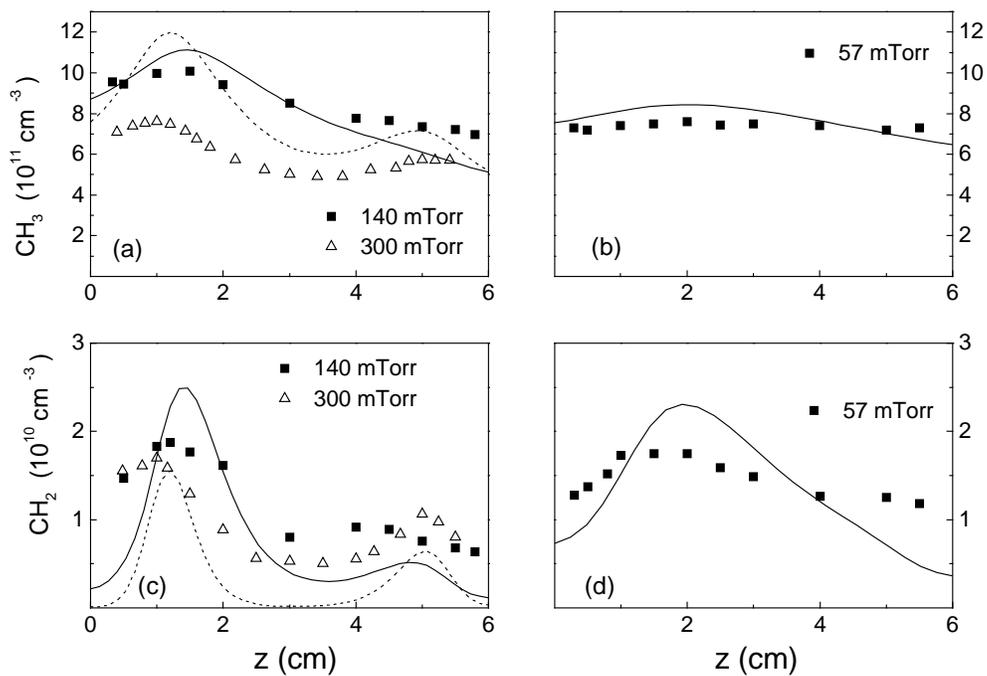}
\caption{Comparison of calculated (curves) and observed in experiment \cite{Sug2}
(points) density profiles for CH$_3$ (a,b) and CH$_2$ (c,d) radicals for
different pressures.
Solid curves and squares correspond for P=140 mTorr (a,c) and 57 mTorr (b,d);
dashed curves and trangles - for P=300 mTorr (a,c).}
\end{figure}

We can say the calculated profiles exhibit the similar behaviour
as the observed in experiment. The shape of described in \cite{Sug2} density
humps and transition to flat profiles with pressure decreasing are 
simulated well. We suppose that the appearance of density humps indicate the 
transition from VD to AS regimes, which is consistent with our simulations.
The quantitative agreement is good in pressure range from 50 to 150 mTorr 
for both radicals, especially for CH$_3$. Also for CH$_3$ it is
within 30\%  in range 50$\div$300 mTorr. As for CH$_2$, the calculated
densities in plasma bulk are underestimated for pressures 
larger than 150 mTorr. 

\section {Summary}

Accelerated PIC-MCC method was applied to two-dimensional simulation of 
capacitively coupled radio frequency discharge in methane.
The obtained spatial distributions of mean electron
energy and electron--methane reaction rates show
existence of two regimes of discharge glow. The first, with active sheaths, is
characterized by hot electrons localized in electrode sheaths and relatively 
cold electrons in other discharge volume, hence the reactions proceed
in electrode sheaths regions and are suppressed in the plasma bulk.
For the case of the second one, with volume domination, the electron
energy is more uniform and the reactions take place in all discharge volume.

The transition between discharge regimes for one--dimensional geometry
was previously studied in \cite{1D} using combined PIC--MCC model. 
Unlike the 1D results, 
where the transition occurs abruptly at some critical current density,  
in 2D case transition is gradual, with continuous evolution of
mean electron energy and reaction rates profiles.
For the considered reactor geometry it was found, that
the VD regime is observed for low rf voltages (40-60 V), 
and turns to AS with voltage
increasing. The transition proceeds in $U_{rf}$
range from 60 to 120 V and is noticed firstly at the spatial
profile of electron--methane reaction rates. For $U_{rf}$ over 120 V the 
regime is definitely AS.
It was also noted, that the mean electron energy profile tends to saturate
for $U_{rf}$ over 150 V, and changes very little with further 
voltage increasing. For another reactor geometry 
the quantitative results may differ.

To find the gas mixture composition in discharge volume,
simulations of chemical reactions in methane plasma were also performed, 
with the diffusion and convection flux of species included.
The results show good agreement with the 
known experimental data, especially for pressure range from 50 to 150 mTorr.
For more wide range to 300 mTorr, a qualitative agreement, 
namely the behaviour of radical density profiles shape, is obtained too.

\section*{Acknowledgments}
This work was supported by NATO grant SfP-974354 "SfP-Diamond deposition".

\end{document}